\documentstyle[aps,multicol,epsfig,prl]{revtex}
\def\C60{C$_{60}$}
\def\Tc{T$_c$ }
\def\O{{\cal O}}

\input epsf

\begin{document}

\title
{Superconductivity in hole-doped \C60 from electronic correlations.}

\author{M.~Granath and S.~{\"O}stlund}
\address
{Institute for Theoretical Physics, Chalmers University of Technology and G\"{o}teborg University, 
S-412 96 G\"{o}teborg, Sweden}

\date{\today}
\maketitle
\begin{abstract}
We derive a model for the highest occupied molecular orbital band of a 
\C60 crystal which includes on-site electron-electron interactions. 
The form of the interactions are based on the icosahedral symmetry of the \C60 molecule together 
with a perturbative treatment of an isolated \C60 molecule. Using this model
we do a mean-field calculation
in two dimensions on the [100] surface of the crystal. Due to the multi-band nature  
we find that electron-electron interactions can have a profound effect on the 
density of states as a function of doping. 
The doping dependence of the transition temperature can then be  
qualitatively different from that expected from simple BCS theory based on the 
density of states from band structure calculations.
\end{abstract}

\begin{multicols}{2}
\narrowtext
Superconductivity in \C60 has generally been ascribed to a phonon 
mechanism due to strong electron-phonon coupling for some \C60 intramolecular
modes \cite{phonons}.  
However, due to their high energy and the narrow electronic band width questions have been 
raised about the effectiveness of retardation for reducing the strong Coulomb repulsion in these 
materials \cite{Anderson}.
In addition, a number of features suggest that these
materials are very exotic, including Mott insulating behavior and  
so called ``bad metal'' behavior with resistivities which do not saturate at high 
temperatures 
\cite{Gunnarsson_review,Steve_new}. Such behavior is reminiscent
of the high-\Tc cuprates where electron correlations are generally accepted to play a 
crucial role.  

Another issue which is not well understood is the variation of \Tc with doping. 
The various alkali-doped materials have \Tc's that are 
maximized near the half-filled LUMO (lowest unoccupied molecular orbital) band, 
i.e. 3 electrons per \C60, and confined to a narrow doping range around this. 
This variation of \Tc with doping does not correspond to the density of
states (DOS) as given by band structure calculations \cite{Gunnarsson_review}. Again, a clear indicator that correlation effects are important.

Here we study the effects of  
electronic correlations on a crystal of \C60 molecules based on the strong intra-molecular 
electron-electron interactions. 
Our approach, which expands on earlier work by Chakravarty et al. \cite{Sudip} (see also \cite{Baskaran}
for similar and independent ideas), can be summarized as 
follows. We solve the Hubbard model on a truncated icosahedron, i.e. a single 
\C60 molecule, in second order perturbation theory in the on-site repulsion
U. We do this for the HOMO (highest occupied molecular orbital) states given by 
diagonalizing the
tight-binding ($U=0$) Hamiltonian. Based on the perturbative spectrum we formulate an effective 
interacting Hamiltonian in terms of holes characterized by orbital angular 
momentum and spin. We then consider a crystal of \C60 molecules with nearest neighbor hopping
and where this effective Hamiltonian for the interactions on a single \C60 molecule correspond to
on-site interactions. Subsequently we do a standard BCS/Hartree-Fock calculation on 
the [100] surface of an fcc crystal using this 
lattice Hamiltonian. The Hartree-Fock calculation on a surface- and hole-doped
\C60 crystal is a model calculation. Nevertheless we feel that the method 
presented as well as the qualitative features of the results are relevant also
to the alkali doped materials. 

We find that \Tc is peaked close to three holes and strongly 
suppressed at five holes where the DOS based on band structure is maximized. 
This striking deviation from the behavior expected from a BCS calculation based on
band structure is
related to a strong renormalization of the DOS due to the interactions. As a signature of the strong 
electron-electron interactions we also find that depending on the details of the 
interactions 
and band structure there may be non-magnetic Mott insulating states at even integer fillings. 
Mott insulating behavior is indeed seen in 
alkali doped compounds with a doping of two or four electrons per molecule 
\cite{Kiefl,Yildrim}.  In addition,
an equivalent analysis for the LUMO band gives a pair-binding interaction which is roughly 60\% 
of that for the HOMO band in the relevant parameter regime, which suggests 
a possibility for higher \Tc's for a hole-doped material.
  
\paragraph*{Perturbation Theory:}
Let us start by considering the following Hubbard model on a single \C60 molecule,
\begin{equation}
H_{C_{60}}=-\sum_{ij,\sigma}t_{ij}(c^{\dagger}_{i\sigma}c_{j\sigma}+\mbox{h.c.})
	+\frac{U}{2}\sum_{i,\sigma}n_{i\sigma}n_{i-\sigma}\,,
\end{equation}
where the only non-vanishing hopping integrals are $t_{ij}=t$ for nearest neighbor (nn) hopping 
on pentagons and 
$t_{ij}=t'$ for nn hopping between pentagons. We use $t=2eV$ and $t'/t=1.2$ 
in accordance with the values used in \cite{Sudip} and allow $U$ to vary. 
Values of $U\sim 5-12 eV$ have been suggested in the literature \cite{Gunnarsson_review}.  

We do standard second order perturbation theory in Hubbard U.
Since the Hamiltonian is spin rotationally invariant the states fall into degenerate 
sets corresponding to irreducible 
representations of the icosahedral group $I_h$ and spin.
The states are well characterized by angular momentum and only weakly 
split by the icosahedral symmetry.

The validity of second order perturbation theory
for the large-U Hubbard model and the neglect of longer range Coulomb interaction for this problem 
has been under debate\cite{Gunnarsson_review}. It has been shown by exact diagonalization that 
for small Hubbard clusters (e.g. the 12-site truncated tetrahedron) the second order perturbation theory is
qualitatively correct giving positive pair-binding energies for moderately large U \cite{White}. 
In addition, longer ranged repulsions are more effectively screened by the metallic environment 
than the high energy second order processes that give rise to the attraction \cite{Lammert}.
%Other examples of attraction from repulsion in finite systems include doped 2 and 3-leg t-J ladders 
%\cite{Jeckelmann} as 
%well as one-dimensional Hubbard rings where the pairing energy is maximized for intermediate size and 
%U \cite{Steve_new}. 
Here, we explore the consequences of this model assuming that it gives a reasonable estimate of the 
molecular spectrum and the resulting pair attraction. 
\paragraph*{Effective Interactions:}  
Using the perturbation theory results we derive a set of interactions for the crystal.
Although the perturbative result contains terms to higher order in fermion operators, we take the 
effective Hamiltonian 
\begin{equation}
H_{eff}=e_0-e_1\sum_{l,\sigma}n_{l\sigma}
	+t_iT^i_{klmn\alpha\beta\gamma\delta}c^{\dagger}_{k\alpha}
	c^{\dagger}_{l\beta}c_{m\gamma}c_{n\delta}\,,\label{Heff}
\end{equation}
which acts on a space of
one particle states in the five-dimensional $H_u$ representation of $I_h$. 
Here $c^{\dagger}_{k\alpha}$ creates a hole with quantum number
$k=1,...,5$ and spin $\alpha$, 
$n_{l\sigma}$ is the number operator and alike indices are henceforth summed over.
The $e_0$, $e_1$
and $t_i$ are parameters and the $T^i_{klmn\alpha\beta\gamma\delta}$ are tensors
chosen such to make the four-fermion terms $T^i$ invariant independently under spin and 
icosahedral symmetry. 

Group theory reveals that
there are nine such independent four-fermion terms. These can 
be derived by constructing all two-fermion terms  
$c_{k\alpha}c_{l\beta}$ transforming in a particular representation of spin and 
angular momentum and taking the tracing with their hermitian conjugates. 
We can write for the product of two fermions in the representation $H$ of $I_h$ and 
spin$-1/2$,
$H\otimes H=(A\oplus G\oplus 2H)_s+(T_1\oplus T_2\oplus G)_a$,
$1/2\otimes 1/2=0_a\oplus 1_s$
where $s$ and $a$ mean the symmetric and antisymmetric parts of the tensor products
and where $A,T_1,T_2,G$ and $H$ are the 1,3,3,4 and 5 dimensional representations of 
$I_h$ respectively.

The product of two
anticommuting fermion operators thus reduces into seven irreducible parts, given by  
finding the antisymmetric part of the product of angular momentum and spin. 
We then construct the invariant four-fermion operators 
$T^i$, with corresponding coupling constants $t_i$, labeled according to from which
two-fermion
operators they are constructed using the composite index 
\begin{eqnarray}
\label{composite_index}
i&=&\{(A,0,0),(H,2,0),(G,4,0),(H,4,0),(T_1,1,1),\\\nonumber
&&(T_2,3,1),(G,3,1)\}\,.
\end{eqnarray} 
Here $(R_i,L_i,S_i)$ indicates icosahedral 
representation, corresponding angular momentum in 
the case of full rotational symmetry, $O(3)$, and spin, respectively. The tensors
$T^i_{klmn\alpha\beta\gamma\delta}$ are normalized such that they are projection operators into the 
$i$th irreducible subspace of the two-fermion products. We neglect the additional two invariants that 
can be constructed from tracing
the two different realizations of $H$ together since these are not allowed under
$O(3)$ and since the deviance from full rotational symmetry is small.

\begin{figure}
\narrowtext
\begin{center}
\leavevmode
\noindent
\centerline{\epsfxsize=3.2in \epsffile{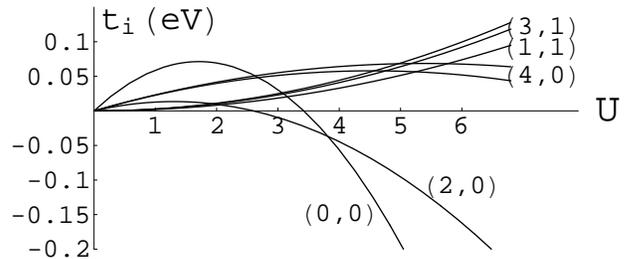}}
\end{center}
\caption{Coupling constants of the effective interactions labeled according to $(L,S)$. The split due
to icosahedral symmetry is not explicitly labeled. ($U$ is in units of $t=2eV$)}
%\protect{}
\label{tsofU}
\end{figure}

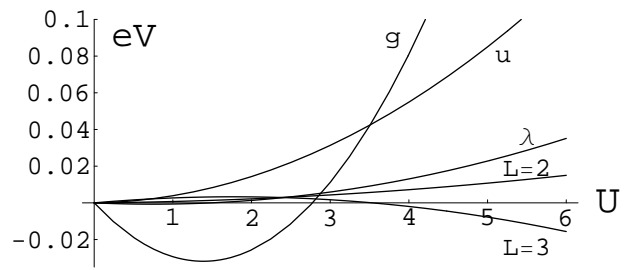
\begin{figure}
\narrowtext
\begin{center}
\leavevmode
\noindent
\centerline{\epsfxsize=3.2in \epsffile{opsofUsmall.epsi}}
\end{center}
\caption{Magnitude of the coupling constants of $\vec{S}^2$, $n^2$, and $\vec{L}^2$ respectively
$g,u,\lambda$,
and the $L=2$ and $L=3$ tensor invariants. ($U$ is in units of $t=2eV$)}
%\protect{}
\label{opsofU}
\end{figure}

The effective Hamiltonian (\ref{Heff}) is then used to match the spectrum found from 
the perturbation theory. $e_0$ is given by the energy of the neutral molecule (filled HOMO),
$e_1$ by the 10-fold degenerate 1-hole states and the 2-hole spectrum is in one-one
correspondence to the seven four-fermion terms with energies $e_0-2e_1+2t_i$, which 
fixes $t_i=t_i(U,U^2)$ as shown in figure \ref{tsofU}.
%\begin{figure}[ht]
In spite of its relative simplicity 
this effective model reproduces the trend found in the 
perturbative calculation, namely that for moderately large $U$ states with low spin and low 
``angular momentum'' 
have lower energy, i.e. Hund's rule is not valid. 
To make this statement more transparent we can consider the conventional four-fermion
operators $n^2$, $\vec{S}^2$ and $\vec{L}^2$, invariant under $O(3)\times SU(2)$. 
To fit these to the  
normal ordered four-fermion operators $T^i$ we also need
the tensor invariants of angular momentum ${\cal Q}^a_L{\cal Q}^a_L$ 
(no sum over $L$) where 
${\cal Q}^a_L=Q^a_{L,ij}c^{\dagger}_{i\alpha}c_{j\alpha}$ transform as the $L=$2,3 or 4 
representation of $SO(3)$ ($a=1,...,2L+1$). We find, 
$n^2=n+\sum_iT^i$,
$\vec{S}^2=\frac{3}{4}n+\sum_i(\frac{S_i(S_i+1)}{2}-\frac{3}{4})T^i$ and 
$\vec{L}^2=6n+\sum_i(\frac{L_i(L_i+1)}{2}-6)T^i$,
with $i$, $S_i$ and $L_i$ as defined in (\ref{composite_index}) and with similar 
expressions for the other operators. Apart from the
small split due to icosahedral symmetry, which we average over,
we get new coupling 
constants as shown in fig. (\ref{opsofU}). (The $L=2,3$ invariants
are normalized as $\vec{L}^2$ and due to overcompleteness we
choose not to include the $L=4$ tensor invariant.) 
We find that for large $U$, $n^2$, $\vec{L}^2$ and $\vec{S}^2$ dominate the energetics and
we will subsequently 
only use these as a minimal model expected to capture the important physics of the 
interactions.  
 
\paragraph*{Lattice Hamiltonian:}
We can now write down the lattice Hamiltonian
\begin{eqnarray}
\label{Hreal}
H&=&\sum_{<\vec{r}\vec{r}'>}t_{\vec{r}\vec{r}',kl}(c^{\dagger}_{\vec{r},k\sigma}
c_{\vec{r}',l\sigma}+\mbox{h.c.})\\\nonumber
&+&\sum_{\vec{r}}g\vec{S}_{\vec{r}}\cdot\vec{S}_{\vec{r}}+
\lambda\vec{L}_{\vec{r}}\cdot\vec{L}_{\vec{r}}
+un^2_{\vec{r}}\,,
\end{eqnarray}
where $\vec{r}$ are the sites of the lattice and $<\vec{r}\vec{r}'>$ runs over the range of 
intermolecular hopping.
This model is quite general and could be used also for electron doped systems in two or three
dimensions \cite{M&S_in_progress}. Here we consider a model where the charge is 
confined to the [100] surface of the fcc crystal, i.e. a two dimensional system 
where $<\vec{r}\vec{r}'>$ correspond to nn on a square lattice. 
From the perturbative calculation 
(fig. \ref{opsofU}) we take $g,\lambda,u>0$. 
Note that the confinement of electron propagation to the surface 
completely breaks the 
five-fold degeneracy of the HOMO states. This will be manifest in the tight-binding 
part of the Hamiltonian which reflects the symmetry group of the surface. 

Typical hopping integrals are of the
order of 0.1eV \cite{Gunnarsson} which is comparable to the interactions 
(fig. \ref{opsofU}), the problem is in an intermediate coupling regime.  
Nevertheless, we do a standard BCS/Hartree-Fock 
construction, replacing the Hamiltonian (\ref{Hreal}) by a reduced
non-interacting Hamiltonian. We keep only spatially 
uniform superconducting mean-fields 
$b_{kl\alpha\beta}=\frac{1}{V}\sum_{\vec{r}}<c_{\vec{r},k\alpha}c_{\vec{r},l\beta}>$
and mean-fields of the number operators 
$n_{l\alpha}=\frac{1}{V}\sum_{\vec{r}}<c^{\dagger}_{\vec{r},l\alpha}c_{\vec{r},l\alpha}>$. 
We can then derive the 
following effective Hamiltonian in momentum ($p$) space 
\begin{eqnarray}
\label{H_red}
\nonumber
&H_{MF}&=H_0+H_{pair}+H_{HF}
=\sum_p (t(p)_{kl}-\mu\delta_{kl})c^{\dagger}_{p,k\sigma}c_{p,l\sigma}\\
&-&\sum_{L,S}(c^{\dagger}_{p,l\beta}c^{\dagger}_{-p,k\alpha}\O^{L,S}_{kl\alpha\beta}
+\mbox{h.c.})+h_{HF,kl}c^{\dagger}_{p,k\alpha}c_{p,l\alpha}\,,
\end{eqnarray} 
where we included a chemical potential $\mu$. We define the components of the order parameter
\begin{equation}
\label{deltas}
\O^{L,S}_{kl\alpha\beta}=V_{L,S}\sum_{i:L_i=L,S_i=S}T^i_{klmn\alpha\beta\gamma\delta}
b_{mn\gamma\delta}\,,
\end{equation}
where $V_{L,S}=(\frac{3}{4}-\frac{S(S+1)}{2})g+(6-\frac{L(L+1)}{2})\lambda-u$ and 
$V_{L,S}>0$ corresponds to attraction. 

Assuming no net magnetization the Hartee-Fock terms can be 
written (no sum over $l$) 
\begin{equation}
\label{H_HF}
h_{HF,ll}=-\frac{3}{4}gn_l-\lambda\tau_l+u(n_l+2\sum_{k\neq l}n_k)\,,
\end{equation}
where $n_l=n_{l\uparrow}+n_{l\downarrow}$ is the total particle number with angular momentum
component $l$ and $\vec{\tau}=(3n_4+3n_5,4 n_3+n_4+n_5,4n_2+n_4+n_5,
3n_1+n_2+n_3+n_5,3n_1+n_2+n_3+n_4)$.
(In addition there is one off-diagonal component 
$\sqrt{3}\lambda(n_5-n_4)(c^{\dagger}_{p,1\alpha}
c_{p,2\alpha}+\mbox{h.c.})$, which is included in our calculations but which will in general be 
small.) 

The tight-binding part of the Hamiltonian for the simplest (unidirectional) crystal structure takes the form
\begin{equation}
\label{H_TB}
t(p)_{kl}\doteq\left [\matrix{t_1f_p&t_{12}f_p&t_{13}g_p&0&0\cr
           t_{12}f_p&t_2f_p&t_{23}g_p&0&0\cr
           t_{13}g_p&t_{23}g_p&t_3f_p&0&0\cr
           0&0&0&t_4f_p&t_{45}g_p\cr
           0&0&0&t_{45}g_p&t_5f_p}\right]
\end{equation} 
where $f_p=\cos(p_x)\cos(p_y)$ and $g_p=\sin(p_x)\sin(p_y)$. 

We have taken hopping parameters 
from \cite{Gunnarsson}. For hole hopping we have, 
$t_1=-.107,t_2=.198,t_3=.134,t_4=-.032,t_5=-.170,t_{12}=.087,
t_{13}=.073,t_{23}=.162,t_{45}=.115$ eV. The Hamiltonian has the symmetry
$t(p)=-t(p+(\pi,0))$ implying a symmetric band structure around zero energy where all bands
will be half-filled and there are Van-Hove singularities at zero energy at 
$(p_x,p_y)=(\pm\pi/2,\pm\pi/2)$. 

The Hartree-Fock terms (\ref{H_HF}) have interesting 
properties related to their multiband nature.  
For positive $g$ and $u$ it is
energetically favorable to fill up as few bands as possible for a given particle number. The term
$\vec{S}_{\vec{r}}\cdot\vec{S}_{\vec{r}}$ gives
on-site spin-triplet states with higher energy than
singlets, so that by putting particles in a single angular momentum state the energy can be 
lowered by exclusion, and similarly for the on-site charging energy $n^2_{\vec{r}}$, 
where two particles with the same spin and angular momentum cannot occupy the same site. The 
$\vec{L}_{\vec{r}}\cdot\vec{L}_{\vec{r}}$ term on the other hand gives rise to an 
anisotropic attraction between the components.

For positive parameters $g,u$ and $\lambda$ the $L=0$, $S=0$ pairing channel of (\ref{H_red}) is 
strongest and we can
expect this to dominate. But, since the rotational invariance of the \C60 molecules is broken by 
the lattice, subdominant order parameters with non-zero angular
momentum appear and in general all three $S=0$ order parameters may be 
non-zero.

What kind of physics can we expect from this model?
For large $g$ or $u$ there
may be Mott insulating states at even integer $2n$ filling where $n$ bands of angular momentum 
states will fill up completely. If $u$ is not very large compared to $g$ the insulating state will
be non-magnetic due to the low energy of molecular singlets. (For the regime $u\gg g$, not realized here, 
there may also be magnetic insulating states at odd integer filling.)
The pairing terms compete with a putative insulating 
state due to the Hartree-Fock terms so that even for large $g$ or $\lambda$  
there may be 
s.c. ground states also at even integer fillings, although a suppression 
of \Tc is likely due to the low
DOS when the bands are nearly filled or empty. In general we can expect the Hartree-Fock
terms to completely recast the DOS compared to that given by the band structure and consequently 
also T$_c$'s. 

\begin{figure}
\narrowtext
\begin{center}
\leavevmode
\noindent
%\centerline{\epsfxsize=3.2in \epsffile{dofn2.epsi}}
\centerline{\epsfxsize=3.2in \epsffile{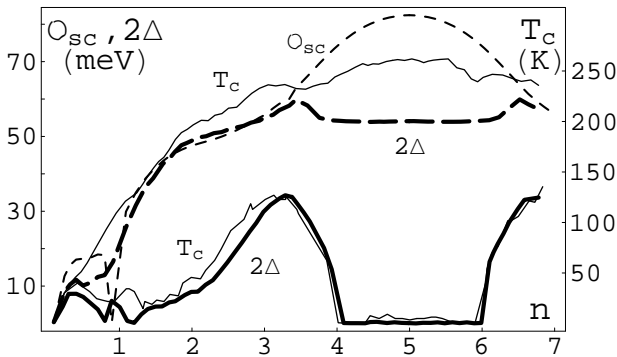}}
\end{center}
\caption{Spectral gap $2\Delta$ and norm of $\O_{sc}$ at T=0 and 
\Tc for parameters $u=.09,g=.06,\lambda=.02$ eV as a function of doped holes. The lower (upper) curves are 
with (without) Hartree-Fock terms.}
%\protect{}
\label{dofn}
\end{figure}

\paragraph*{Results:}
By numerical iteration of the mean particle number in the five angular
momentum components and 
the s.c. mean-fields at fixed chemical potential we arrive at self-consistent 
solutions. For all plots the 
system size is $100\times 100$ with at least ten sampling points per unit shift in particle 
number.  
Figure (\ref{dofn}) shows the energy gap $2\Delta$, the norm of the s.c order parameter 
(defined as $\sqrt{Tr\O\O^{\dagger}}$ with $\O=\sum_{L,S}\O^{L,S}$)
at T=0 and \Tc as a function of doped holes (up to 7 holes) for parameters $u=.09,g=.06,\lambda=.02$ eV, 
both with  and without Hartree-Fock terms. We find that \Tc scales roughly linearly with 
$\Delta$ and the reduced gap $\frac{2\Delta}{T_c}\approx 3.2$ is close to the 
weak coupling BCS value (3.53). For the calculation with H-F terms the magnitude of the order 
parameter (not in the figure) fits very well with $2\Delta$. Without H-F terms 
there is a deviation from this fit around 5 holes, due to a momentum dependent gap.

\begin{figure}
\narrowtext
\begin{center}
\leavevmode
\noindent
\centerline{\epsfxsize=3.2in \epsffile{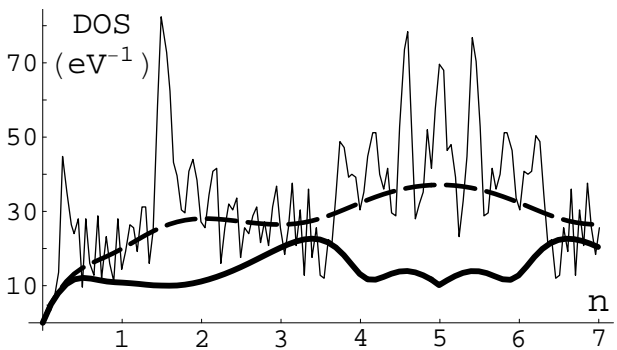}}
\end{center}
\caption{Density of states. The thick solid line is for parameters as in fig. (\ref{dofn}), 
calculated at 150K. The thin (dashed) line is the DOS from band structure at T=0 (T=150K).}
%\protect{}
\label{dosofn}
\end{figure}

Fig. \ref{dosofn} shows the density of states, 
$\frac{\partial n}{\partial\mu}$, for the same parameters and with Hartree-Fock terms 
and the DOS from the noninteracting band structure (\ref{H_TB}), i.e. the DOS without H-F terms. 
Since we find self-consistent solutions of both the band fillings and the gaps we 
calculate the DOS at finite temperature, above \Tc, for the realization with H-F terms. 

The values of $\lambda$ and $u$ chosen here correspond roughly to the perturbative spectrum 
(fig. \ref{opsofU}) at $U=5$, but we have reduced $g$ significantly 
in order for the pairing attraction not to dominate completely. 
One could argue that the second order perturbation theory can be 
expected to overestimate the core polarization effect that gives the 
$\vec{S}^2$ term.
Of course, the actual magnitude of \Tc and the gap that we find should not be taken too literally 
since parameters are only rough estimates and we are doing mean-field theory at relatively 
strong coupling. The important result is the qualitative behavior of \Tc as a function of
doping.  

Without Hartree-Fock terms, \Tc follows the DOS of the band structure with a corresponding 
maximum at the half-filled band.
This should be contrasted with the results for the full Hamiltonian which has a \Tc that
is maximized close to three holes and which is strongly suppressed at five holes. 
The DOS is suppressed at four and six holes 
due to the commensurate lock-in discussed above. In fact, the influence of these special fillings 
is such that the DOS is low in the whole region 
between four and six holes where it is the highest without interactions.

\thanks{
We thank S.A. Kivelson for valuable comments. This research was supported by STINT and 
the Swedish Research Council.}

\end{multicols}
\end{document}